\documentclass[authoryear,preprint]{elsarticle}
\usepackage{pifont}
\usepackage{natbib}
\usepackage{geometry}
\usepackage{graphicx}
\usepackage{amssymb}
\usepackage{amsmath}
\usepackage{amsthm} 
\usepackage{textcomp}
\usepackage{mathabx}
\usepackage{color}
\usepackage{url}
\journal{Icarus}

\begin{document}

\begin{frontmatter}

\title{Asteroid 4 Vesta: dynamical and collisional evolution during the Late Heavy Bombardment}

\author[sim]{S. Pirani}
\ead{simona@astro.lu.se}

\author[label1,label2]{D. Turrini}
\ead{diego.turrini@iaps.inaf.it}

\address[sim]{Lund Observatory - Department of Astronomy and Theoretical Physics - Lund University - S\"olvegatan 27 - 22100 Lund - Sweden}
\address[label1]{Istituto di Astrofisica e Planetologia Spaziali INAF-IAPS - Via Fosso
del Cavaliere, 100 - 00133 Roma - Italy}
\address[label2]{Departamento de Fisica - Universidad de Atacama - Copayapu 485 - Copiap\'o - Chile}

\begin{abstract}

Asteroid 4 Vesta is the only currently identified asteroid for which we possess samples in the form of meteorites. These meteorites revealed us that Vesta is a differentiated body and that its differentiation produced a relatively thin basaltic crust that survived intact over its entire collisional history. The survival of the vestan basaltic crust has long been identified as a pivotal constraint in the study of the evolution of the asteroid belt and the Solar System but, while we possess a reasonably good picture of the effects of the last 4 Ga on such a crust, little is know about the effects of earlier events like the Late Heavy Bombardment. In this work we address this gap in our knowledge by simulating the Late Heavy Bombardment on Vesta in the different dynamical scenarios proposed for the migration of the giant planets in the broad framework of the Nice Model. The results of the simulations allowed us to assess the collisional history of the asteroid during the Late Heavy Bombardment in terms of produced crater population, surface saturation, mass loss and mass gain of Vesta and number of energetic or catastrophic impacts.
Our results reveal that planet-planet scattering is a dynamically favourable migration mechanism for the survival of Vesta and its crust. The number of impacts of asteroids larger than about 1 km in diameter estimated as due to the LHB is $31\pm5$, i.e. about 5 times larger than the number of impacts that would have occurred in an unperturbed main belt in the same time interval. The contribution of a possible extended belt to the collisional evolution of Vesta during the LHB is quite limited and can be quantified in $2\pm1$ impacts of asteroids with diameter greater than or equal to 1 km. The chance of energetic and catastrophic impacts is less than 10\% and is compatible with the absence of giant craters dated back to 4 Ga ago and with the survival of the asteroid during the Late Heavy Bombardment. The mass loss caused by the bombardment translates in the erosion of $3-5$ meters of the crust, consistently with the global survival of the basaltic crust of Vesta confirmed by the Dawn mission. Our analysis revealed that the contribution of the LHB to the cratering of Vesta' surface is not significant and is actually erased by the crater population produced by the following 4 Ga of collisional evolution of the asteroid, in agreement with the data provided by the Dawn mission.
\end{abstract}


\end{frontmatter}

\section{Introduction}

\label{}

Vesta is the only identified, and supposedly intact, parent body of meteorites in the asteroid belt. The HEDs, the Howardite-Eucrite-Diogenite class of meteorites, represent fragments of the surface of the asteroid \citep{mccord70,consolmagno77,gaffey97,desanctis12,prettyman12} and, as such, provide information about their differentiated parent body's internal structure and history. The oldest HEDs samples date the formation of the crust of Vesta to the first 1-2 Ma after the condensation of the first solids of the Solar System \citep{bizzarro05,schiller11} and the complete solidification of the vestan crust and mantle to the first 100-150 Ma \citep{mcsween11}.

Combining the information provided by the HEDs with that supplied in recent years by the NASA mission Dawn, we now know that Vesta and its crust survived intact \citep{russell12,desanctis12,prettyman12,consolmagno15} their crossing of the entire evolution of the Solar System, from catastrophic events such as the Jovian Early Bombardment \citep{turrini14a,turrini14b} and the Late Heavy Bombardment \citep{gomes05,Marchi13} to the decreasing impact flux of the last 4 Ga \citep{turrini14,obrien14,schmedemann14}.

For what it concerns the Late Heavy Bombardment in particular, while the HED show signs of the passage of Vesta through the bombardment in their Ar-Ar ages \citep{Marchi13}, there are currently no estimates of how many impacts this event could have caused, partly due to the fact that the ages of the oldest craters present on the vestan surface are still debated \citep{obrien14,schmedemann14}.

One of the possible causes of the Late Heavy Bombardment is the migration of the giant planets as hypothesized in the Nice Model \citep{tsiganis05,morbidelli05,gomes05,morbidelli07,Levison11}. The resulting shift of orbital resonances through the asteroid belt would have lead to a significant loss of mass \citep{gomes05}, which the most recent studies estimated to have been of the order of a factor of two \citep{minton09,morbidelli10}.

Recent dynamical studies reported that dynamical diffusion into the network of orbital resonances that crosses the asteroid belt should have caused a decrease in the asteroid population from the end of the LHB to now by a factor of two \citep{minton10}. Once this decrease is summed to that due to the LHB itself, it indicates that the pre-LHB asteroid belt should have been about four times more populous than the present one.
Since the current asteroid belt ($\sim2.1-3.3$ au) is estimated to possess about $1.4 \pm 0.5 \times 10^6$ bodies with diameter greater than about $1$ km \citep{bottke05,Jedicke02}, the pre-LHB belt should have possessed a population of about $ 6 \times 10^6 $ asteroids in the same size range. 
Bottke et al. (2012) suggested the possible existence, before the LHB, of an extended asteroid belt comprised between 1.7 and 2.1 au with a mass equal to 16\% of that of the asteroid population comprised between 2.1 and 3.3 au. As a consequence, the initial population of the pre-LHB asteroid belt could have been of about $7\times10^6$ asteroids of diameter greater then 1 km, of which about $10^6$ inhabiting the now depopulated extended belt.

In this paper we investigate, by means of numerical simulations, the orbital and collisional evolution of Vesta during the Late Heavy Bombardment to assess the implications of this event for the survival of the asteroid and of its crust. In our simulations we considered two different dynamical scenarios: the slower migration of the giant planets proposed in \citet{minton09} as resulting from their interaction with a massive planetesimal disk, and the faster migration discussed in \citet{morbidelli10} that is associated to a planet-planet scattering event.

\section{Description of simulations}

This section is dedicated to the description of the methods used for studying the evolution of the asteroid belt and Vesta and the techniques we used to obtain the number and sizes of the impactors on Vesta, the consequences of the impacts for its surface and the mass gained and lost by Vesta due to impacts. 

\subsection{Solar System model}

The model Solar System we considered in this study is composed of the Sun, the asteroid belt, and the giant planets Jupiter and Saturn. The main belt is composed by Vesta (treated as a massive body like Jupiter and Saturn) and a swarm of massless particles.

The present day main belt is a region bounded by a secular resonance with Saturn and a mean-motion resonance with Jupiter, respectively the $\nu_6$ at $\sim 2$ au and the 2:1 at $\sim 3.3$ au from the Sun. If Jupiter and Saturn were on different orbits, also the position of these resonances were different and the pre-LHB population of the asteroid belt would have been distributed differently than the present one. As mentioned above, moreover, it was suggested that the pre-LHB asteroid belt could have extended inward down to semimajor axes of about 1.7 au, with the bulk of the bodies populating this extended belt residing at semimajor axes greater or equal to 1.8 au \citep{bottke12}. In order to assess the role of these now-depleted regions of the asteroid belt we followed \citet{morbidelli10} and considered the pre-LHB belt to extend between 1.8 au and 4.0 au. The semimajor axes of the massless particles we used to simulate the asteroids were extracted through a Monte Carlo method assuming a uniform distribution between 1.8 and 4.0 au. Following \citet{morbidelli10}, primordial eccentricities were uniformly extracted in a range from 0.0 to 0.3 and primordial inclinations in the range from $0^\degree$ to $20^\degree$, in agreement with the present day eccentricities and inclinations of the main belt. The argument of pericentre ($g$), the longitude of the ascending node ($n$) and the mean anomaly ($M$) were randomly extracted from a uniform distribution between 0\textdegree and 360\textdegree. We used a total of $1.6\times10^5$ massless particles divided into 32 annular regions, each region having width $\Delta a=0.06875$ au and being populated by $5000$ test particles (see subsection \ref{abm} for further details on the modeling of the asteroid belt). In order to limit the duration of the simulations we adopted a data-parallel approach to study the two migration scenarios under investigation. Each scenario was investigated through a set of 32 simulations, each including the Sun, Vesta, Jupiter and Saturn and one of the annular regions populated by test particles mentioned above. The initial conditions of Vesta, Jupiter and Saturn were identical in all 32 simulations, as each of them was actually reproducing the evolution of a slice of our model Solar System. At the end of each set of simulations, the outputs of the different simulations was aggregated before performing the analysis.

The initial conditions chosen for Vesta are the proper elements reported on the Asteroid Dynamic Site (\url{http://hamilton.dm.unipi.it/astdys/}) and are shown in Table \ref{tab:Ke1}. 

\begin{table}[!h]
\begin{center}
\begin{tabular}{|l|c|c|c|}
\hline
\hline  & \textbf{VESTA} &   \\
\hline
\hline  & Value &  Units \\
\hline $a_V$  & 2.36152 &  au \\
\hline $e_V$  & 0.0987 &  \\
\hline $i_V$ & 6.356 &  deg \\
\hline
\hline
\end{tabular}
\caption[Proper orbital elements of Vesta]{Proper orbital elements of Vesta}
\label{tab:Ke1}
\end{center}
\end{table}

\begin{table}[!h]
\begin{center}
\begin{tabular}{|l|c|c|c|c|c|c|}
\hline
\hline  & \multicolumn{3}{|c|}{\textbf{JUPITER}}  &  \multicolumn{3}{|c|}{\textbf{SATURN}} \\
\hline
\hline  Scenario & $a_J$ (au) &  $e_J$ & $i_J$ (deg) & $a_S$ (au)&  $e_S$ & $i_S$(deg)\\
\hline  \footnotesize{\citet{minton09}} & $5.40$ & $0.0365$ & $1.5447$ & $8.78$ & $0.0769$ & $2.1176$\\
\hline \footnotesize{\citet{morbidelli10}} & $5.40$& $0.0037$ &$0.0015$&$8.78$ & $0.0077$ & $0.0021$\\
\hline
\hline
\end{tabular}
\caption[Starting Keplerian elements of Giant Planets]{Starting Keplerian elements of Giant Planets}
\label{tab:Ke2}
\end{center}
\end{table}

For the orbital elements of Jupiter and Saturn, we adopted in both considered dynamical scenarios the initial semimajor axes proposed by the Nice Model \citep{tsiganis05,morbidelli07,Levison11}. For the pre-LHB eccentricity and inclination values of the giant planets, we proceeded as follows. In the case of the \citet{minton09} scenario we followed these authors and used the current values of the giant planets eccentricies and inclinations. In the case of the \citet{morbidelli10} scenario, we followed \citet{tsiganis05} and assumed the orbits of the giant planets as almost circular and coplanar. As shown in Table \ref{tab:Ke2}, the initial eccentricities were assumed to be one order of magnitude smaller than the actual ones and the initial inclinations about three orders of magnitude smaller. As a result, both the initial eccentricities and inclinations of the giant planets were of the order of $10^{-3}$.

\subsection{Asteroid Belt model}\label{abm}

In order to characterize the collisional evolution of Vesta across the migration of the giant planets and the Late Heavy Bombardment we need to know the total number of asteroids existing at the time and their mass distribution.

\begin{figure}[!h]
\begin{center}
\includegraphics[width=3in]{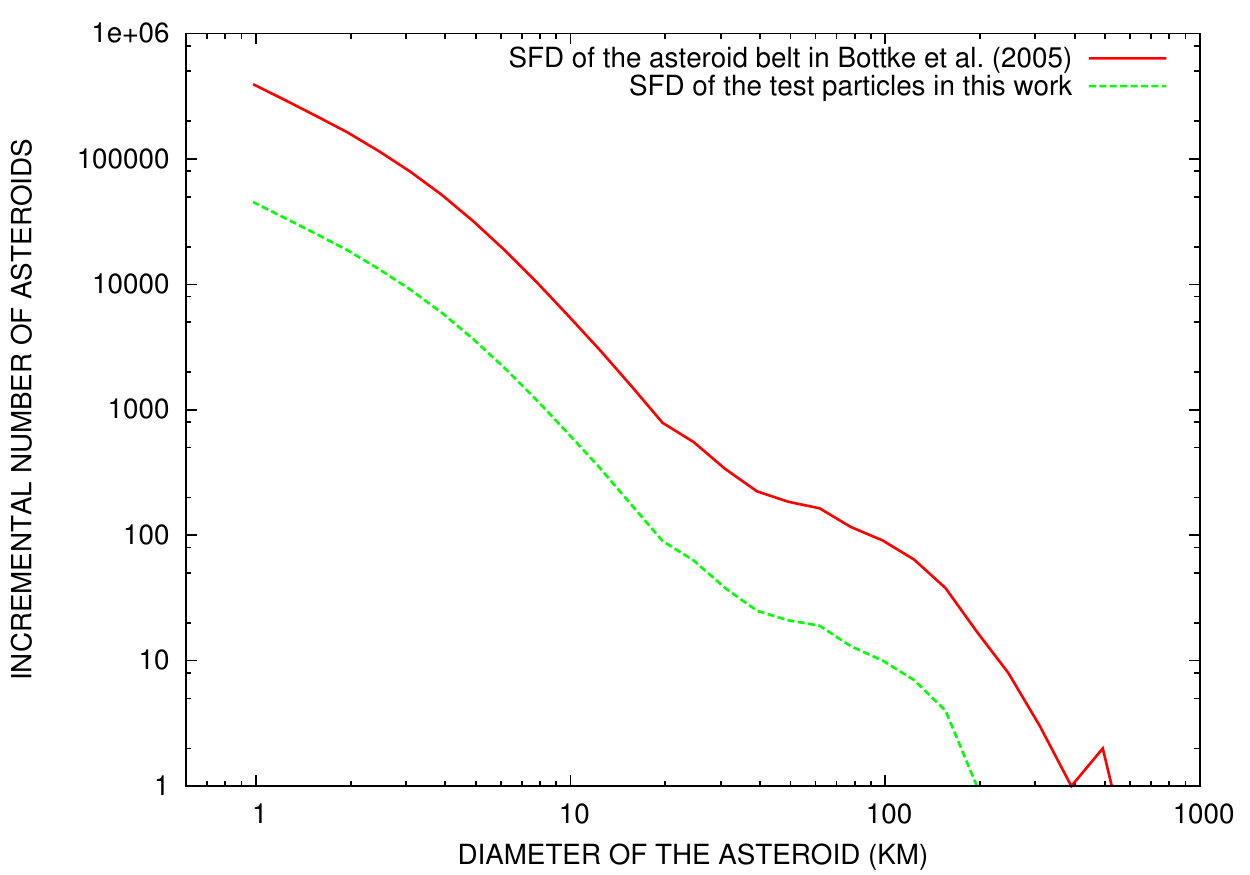}
\caption[In red the main-belt asteroid size distribution reported in \citet{bottke05}. In green the size distribution of the massless bodies used in our simulations.]{\itshape In red the main-belt asteroid size distribution reported in \citet{bottke05}. In green the size distribution of the massless bodies used in our simulations.}
\label{fig:bottke.pdf}
\end{center}
\end{figure}

\citet{bottke05} showed that the size-frequency distribution of the present asteroid belt is stable against collisional evolution over the life of the Solar System. This means that, while the number of asteroids changed with time, the relative number of asteroids of different sizes was mostly constant across the last 4-4.5 Ga. As a consequence, we can use the present size-frequency distribution to associate a mass value to the test particles and assess their effects when impacting on Vesta, as we will describe in the following. The size-frequency distribution we used is based on the one reported by \citet{bottke05} and shown in Fig. \ref{fig:bottke.pdf}.

Concerning the pre-LHB population of asteroids, \citet{minton09} and \citet{morbidelli10} showed that the migration of the giant planets and the Late Heavy Bombardment caused a factor of two depletion relative to the pre-migration belt. Moreover, \citet{minton10} showed that the main belt lost another factor of two of its population from the Late Heavy Bombardment to now. Cumulatively, these results imply that the pre-LHB asteroid population was four times larger than the present one. As the present population of asteroids larger than 1 km in diameter is about $1.36\times 10^6$ \citep{bottke05}, our pre-LHB main belt was populated by $N_{main\,belt}\sim 6\times 10^6$ asteroid with $d\geq 1 km.$ 

Following \citet{morbidelli10} and \citet{bottke12}, we incremented this pre-LHB population by a factor $16\%$ to account for those bodies residing in the now depleted extended belt ranging between 1.7 au and 2.1 au ($N_{e-belt}(d\geq 1 km)\sim 1\times 10^6$). As a consequence, the population of pre-LHB asteroids in our simulations is assumed to have been $N_{pre-LHB}(d\geq 1 km)=N_{main\,belt} + N_{e-belt} \sim 7\times 10^6$. Since we are simulating this pre-LHB main belt with $N_{tot} = 1.6\times 10^5$ particles, each particle represent a swarm of pre-LHB asteroids. The number $\alpha$ of asteroids populating each swarm can be derived by the ratio:

\begin{equation}\label{alpha}
\alpha = \frac{N_{pre-LHB}}{N_{tot}} = \frac{7.0\times 10^6}{1.6\times 10^5} = 43.75
\end{equation}

This ratio is about a factor of $4$ larger than the ratio of the two SFDs shown in Fig. \ref{fig:bottke.pdf}, where we showed the difference between the SFD of our test particles and that of the present asteroid belt (\citealt{bottke05}, therefore without the inclusion of the extended belt suggested by \citealt{bottke12}).

\subsection{Migration models}

We simulated the dynamical evolution of the asteroid belt with the Mercury N-body code \citep{chambers99}, using its hybrid symplectic integrator. Symplectic integrators are faster than conventional N-Body algorithms by about one order of magnitude \citep{wisdom91}, a feature that is particularly important in the framework of this study as we had to simulate the dynamical evolution of the Solar System over timescales of tens of millions years. Moreover, symplectic integrators have the benefit to show no long-term accumulation of errors on the energy. The hybrid symplectic integrator \citep{chambers99} we selected is a refinement of the basic theory of symplectic integrators treated by \citet{wisdom91} and the separable potential method of \citet{duncan98}.
The basic idea of hybrid symplectic integrators is to split the Hamiltonian $H$ of the N-Body system into two or more parts, each one solvable either analytically or by finding an efficient way to integrate it numerically. 

We modified the MERCURY code introducing the migration of the giant planets suggested to be responsible for the Late Heavy Bombardment by \citet{gomes05}. During the migration, the semi-major axes of Jupiter and Saturn evolve across each timestep following the exponential law provided by \citet{minton09}:

\begin{equation}\label{eqn:Minton}
a(t)=a_0+\Delta a[1-exp(-t/\tau)]
\end{equation}

where $a_0$ is the initial semimajor axis, $\Delta a$ is the final displacement, and $\tau$ is the migration e-folding time. When simulating the \citet{minton09} scenario, we adopted their e-folding time of $\tau=0.5$ Ma. When simulating the \citet{morbidelli10} scenario we adopted the value of $\tau=5.0$ Ka suggested by the authors. 
In the case of \citet{morbidelli10} scenario we adopted the analogous laws to increase the values of eccentricity and inclination to the current ones:

\begin{equation}\label{eqn2}
e(t)=e_0+\Delta e[1-exp(-t/\tau)]
\end{equation}

\begin{equation}\label{eqn3}
i(t)=i_0+\Delta i[1-exp(-t/\tau)]
\end{equation}

where $e_0$, $i_0$ are the initial eccentricity and inclination ($\sim 10^{-3}$) and $\Delta e$,$\Delta i$ are the differences with the current values. Before proceeding we must stress that, as pointed out by \citet{morbidelli10}, the real dynamical evolution of the giant planets during a planet-planet scattering event does not follow the smooth variation of the orbital elements described by Eqs. \ref{eqn:Minton}, \ref{eqn2} and \ref{eqn3}. Because of our choice of the migration timescale, however, the dynamical evolution of Jupiter and Saturn in our simulations should provide a reasonable first-order approximation of the dynamical and collisional evolution of the asteroid belt in response to a real planet-planet scattering event.

In both scenarios, we let the system composed of the Sun, Jupiter, Saturn, Vesta and the pre-LHB main belt evolve for a time equal to $5$ Ma $+$ $ 20\tau+5$ Ma. Across the first 5 Ma Jupiter and Saturn do not migrate but they orbit the Sun on their initial orbits. In this period, particles that are dynamically unstable in the pre-LHB configuration of the Solar System are ejected and their possible collisions with Vesta are not considered in estimating Vesta's impact history during the LHB. After this 5 Ma the giant planets begin to migrate. The migration of Jupiter and Saturn lasts for $20$ e-folding times $\tau$, i.e. $10$ Ma in the case of \citet{minton09} scenario and $10^{5}$ years in the case of \citet{morbidelli10} scenario. Lastly, we let the system evolve for another 5 Ma to remove particles that are unstable in the new configuration. Therefore the entire simulations last for 20 Ma in the \citet{minton09} scenario and 10.1 Ma in the \citet{morbidelli10} scenario.
In all simulations, we removed those particles whose semi-major axis became larger than 20 au or smaller than 1.5 au. The first removal distance was imposed to mimic the effect of the icy giants Uranus and Neptune in eliminating those asteroids that cross their orbits, while the former to mimic those of Mars and the terrestrial planets. As pointed out by \citet{morbidelli10}, the removal timescale of Mars is of the order of 100 Ma, therefore far longer than the timespan covered by our simulations. However, as the events simulated in our study are located $\sim$600 Ma after the formation of the terrestrial planets, the adoption of this removal distance allowed for the efficient removal of particles that would have not survived until the LHB in the real Solar System and for avoiding to pollute the results with spurious impact events.

Following \citet{duncan98}, in order to correctly reproduce the dynamical evolution of the simulated bodies we adopted a time-step which is about $1/20$ or less of the orbital period of the innermost body. As the minimum semi-major axis of the disk that we used in the simulations is 1.8 au we selected a time-step of 44 days.

\begin{figure}[!h]
\centering
\centering
\includegraphics[width=2.8in]{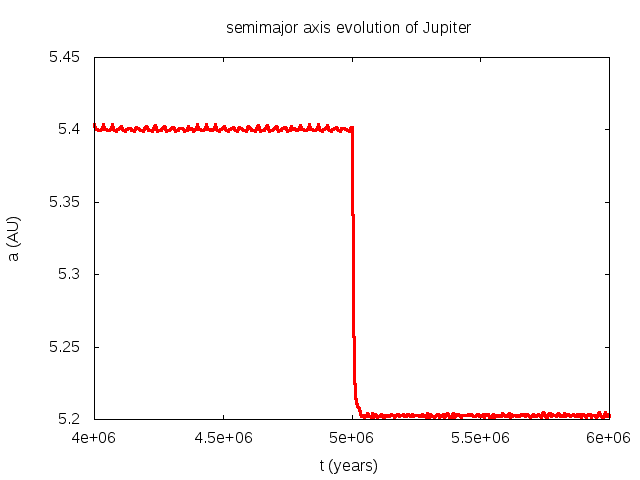}\qquad\includegraphics[width=2.8in]{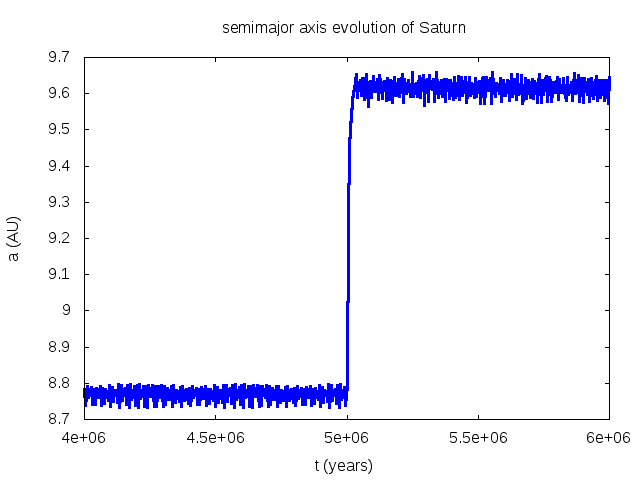}
\includegraphics[width=2.8in]{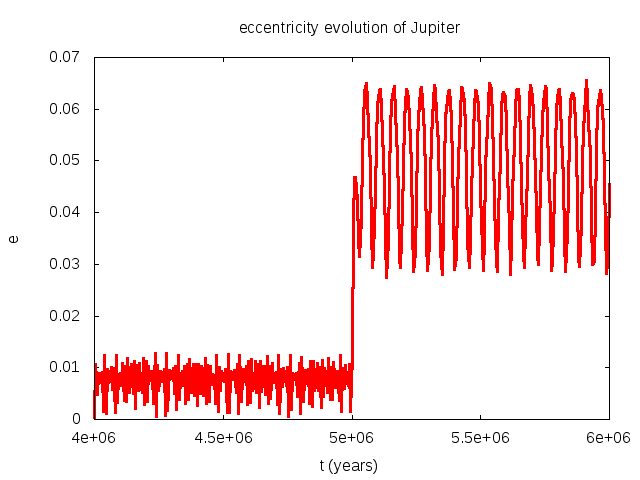}\qquad\includegraphics[width=2.8in]{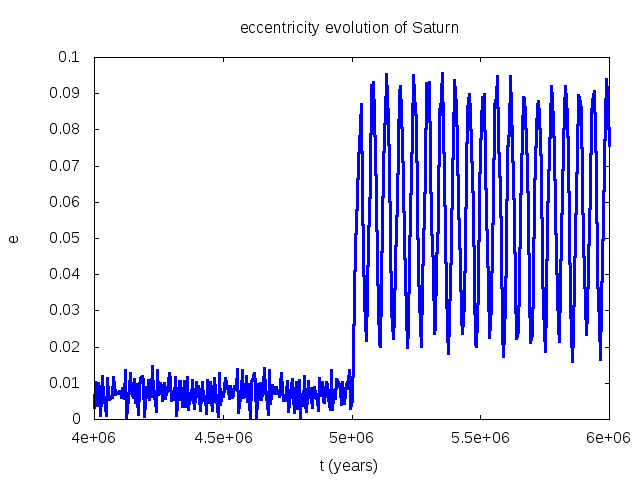}
\includegraphics[width=2.8in]{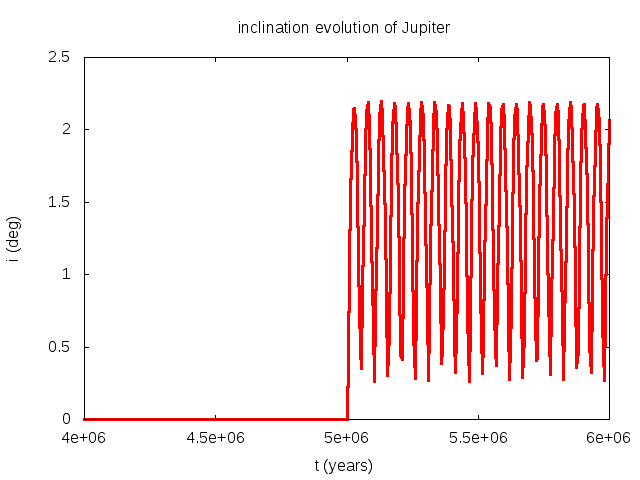}\qquad\includegraphics[width=2.8in]{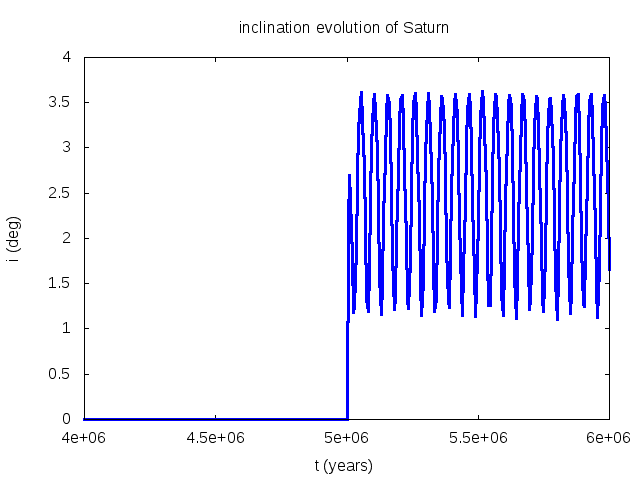}
\caption[Jupiter's and Saturn's dynamical behavior before, across and after the migration in the \citet{morbidelli10} scenario as approximated by Eqs. \ref{eqn:Minton}, \ref{eqn2} and \ref{eqn3}.]{\itshape{Jupiter's and Saturn's dynamical behavior before, across and after the migration in the \citet{morbidelli10} scenario as approximated by Eqs. \ref{eqn:Minton}, \ref{eqn2} and \ref{eqn3}.}}
\label{fig:semi}
\end{figure}

In Fig. \ref{fig:semi} we show an example of the temporal evolution of the semi-major axes, eccentricities and inclinations of Jupiter and Saturn across our simulations as approximated by Eqs. \ref{eqn:Minton}, \ref{eqn2} and \ref{eqn3} in the \citet{morbidelli10} scenario.

\subsection{Collision probabilities}\label{Collision probabilities}

In order to assess the collisional probability of the pre-LHB asteroids on Vesta, we applied the method developed by \citet{turrini11}, i.e. we used a statistical approach based on solving the ray-torus intersection problem between the torus built on the osculating orbit of Vesta and the linearised path of a massless particle across a timestep. 
The basic idea is that the probability that both the impacting body and Vesta will occupy the same spatial region at the same time is the collision probability that we search. This probability can be evaluated as the ratio between the effective collisional time ($T_{eff}$) and the orbital period of the asteroid 4 Vesta ($T_V$). Following \citet{turrini11}, the effective collisional time is the amount of time available for the  collision and it is evaluated as the minimum between the time spent by Vesta ($\tau_V$) and the time spent by impacting particle ($\tau_P$) into the crossed region of the torus.
So the impact probability is given by:

\begin{equation}
 P_{coll}=\frac{T_{eff}}{T_V}=\frac{min(\tau_P, \tau_V)}{T_V}
\end{equation}

The two crossing times, $\tau_V$ and $\tau_P$, are evaluated solving the ray-torus intersection problem.

Once derived the coordinates of the intersection points $i_1$ and $i_2$ between the ray and the torus, we can find the length of the path of the impactor body $d_P$ through the torus and the crossing time $\tau_P$ which, in the linear approximation, is:

\begin{equation}
\tau_P=\frac{d_p}{|v_p|}
\end{equation}

where $|v_p|$ is the modulus of the velocity of the planetesimal since the velocity and the path are parallel. From the intersection points we can also derive the angular width of the crossed section of the torus $\Delta \theta_V$ and the time $\tau_V$ Vesta spends into the crossed region:

\begin{equation}
\tau_A=\Delta \theta_V\frac{T_V}{2\pi}=\frac{\Delta \theta_V}{\omega_V}
\end{equation}

where $\omega_V = n_V = \frac{2\pi}{T_V}$ is the angular velocity (coinciding with the orbital mean motion for circular orbits) of the asteroid.

Once the probability of each event is determined, we can extrapolate the expected number of collisions in the real pre-LHB main belt simply by multiplying the impact probability by the $\alpha$ parameter from Eq. \ref{alpha}. 

\subsection{Impact cratering}

Knowing the impact velocity, we can determinate the diameter of the crater that each impactor creates on the surface of Vesta using the scaling law for basaltic targets described in \citet{holsapple07} in its general form valid across the strength and gravity regimes \citep{holsapple07,turrini14a}: 

\begin{equation}
\frac{R_{crater}}{r_{imp}}=0.93\left(\frac{g r_{imp}}{V_{rel}^2}\right)^{-0.22}\left(\frac{\rho_{imp}}{\rho_V}\right)^{0.31}+0.93\left(\frac{Y}{\rho_V V_{rel}^2}\right)^{-0.275}\left(\frac{\rho_{imp}}{\rho_V}\right)^{0.4}
\end{equation}

where $r_{imp}$, $V_{rel}$ , $\rho_{imp}$ are respectively the radius, impact velocity and the density of the impactor, $\rho_V=3090$ $kg/m^{3}$ and $g=0.25$ m/s$^{2}$ are the surface density and the surface gravity of Vesta (see \citealt{russell12,turrini14a}), and $Y=7.6$ $MPa$ is the strength of the target material (assumed to behave as soft rock to mimic the presence of regolith on Vesta's surface, \citealt{turrini14a}). For the density of the projectiles we considered a constant density of 2 $g/cm^{3}$ \citep{britt02}. The previous and other physical parameters of Vesta are shown in Table \ref{tab:PP}.

\begin{table}[!h]
\begin{center}
\begin{tabular}{|l|c|}
\hline
\hline  Parameter &  Dawn \\
\hline  Major axes (Km) & (286.3/278.6/223.2) $\pm$0.1 \\
\hline  Mean radius (Km) & 262.7 $\pm$0.1 \\
\hline  Volume (Km$^3$) & 74.970 $\times$ 10$^6$\\
\hline  Mass (Kg) & 2.59076 $\pm$ 0.00001 $\times$ 10$^{20}$ \\
\hline  Estimated crustal density (kg m$^{-3}$) & 3090 \\
\hline  Bulk density (kg m$^{-3}$) & 3456 $\pm$ 1\% \\
\hline  Rotation rate (deg/day) & 1617.333119 $\pm$ 0.000003 \\
\hline
\hline
\end{tabular}
\caption[Vesta's physical parameters as measured by Dawn \citep{russell12}]{Vesta's physical parameters as measured by Dawn \citep{russell12}}
\label{tab:PP}
\end{center}
\end{table} 

From the number and the diameter distribution of the craters we can built a plot with the density of the cumulative number of craters in each bin of crater diameter. 
This way we can get information on the contribution of the Late Heavy Bombardment to the saturation of the surface of Vesta. This in turn allows us to estimate how far back in time we can look with the crater record of Vesta and if we can find craters due to pre-LHB events. In order to build the plot, we need to divide the crater record we simulated into diameter bins: following \citet{melosh89}, the minimum diameter we considered is 1 $km$ and each bin goes up to $\sqrt{2}$ times the preceding value, i.e. $D_{i+1}=\sqrt{2}D_{i}$. 

To minimize possible effects due to small number statistics, particularly for the largest bodies, the simulated collisional history is computed by averaging over 1000 Monte Carlo extractions for each collisional event recorded in the simulations (See also section \ref{Mass lost and accreted} for further details).

\subsection{Mass lost and accreted}\label{Mass lost and accreted}

In order to determine the evolution of Vesta's mass during the LHB, we associated a mass to each impacting body by means of a Monte Carlo method based on the size distribution of the present asteroid belt. Fig. \ref{fig:bottke.pdf} gives us the diameter distribution of the asteroids. The Monte Carlo extraction of the mass values is made taking into account the incremental percentage of bodies in every bin of the size-frequency distribution by \citet{bottke05}. That is, we extract a real number in the interval $0.0-100.0$ and use it to select the diameter range of the test particle considered. Once we have the diameter, we can compute the mass for each body considering a constant density of 2 $g/cm^3$ for all the impactors:

\begin{equation}
m_{imp}=\frac{\pi}{6} d^3 \rho_{imp}
\end{equation}

The mass lost by a target body during an impact is a function of relative collisional velocity and both the impactor's and the target's density and physical properties. During the collision, the impactor erodes the surface of Vesta ejecting material from the asteroid. Part of this material is gravitationally recovered by Vesta and part will be lost, which is why the mass loss also depends on the escape velocity of the asteroid. On the other hand, part of the mass of the impactor is accreted by Vesta.

To compute the mass loss $M_{loss}$ suffered by Vesta during the LHB we used the scaling law provided by \citet{holsapple07} for rocky targets (which is independent from the specific cratering regime, i.e. strength- or gravity-dominated) in the angle-averaged form computed by \citet{svetsov11}, where the average is taken weighting over the probability of the different impact angles:

\begin{equation}
M_{loss}=\left[0.03\left(\frac{V_{rel}}{v_{esc}}\right)^{1.65}\left(\frac{\rho_{vesta}}{\rho_{imp}}\right)^{0.2}\right]m_{imp}
\end{equation}\\

To estimate the mass $M_{gain}$ gained by Vesta as a result of the LHB we used the following scaling law for rocky impactors derived by \citet{svetsov11} based on hydrocode simulations focusing on dunite targets and impactors and valid for impact velocities lower than 15 km/s:

\begin{equation}
M_{gain}=(0.14+0.003V_{rel})\ln v_{esc}+0.9V_{rel}^{-0.24}
\end{equation}

It must be noted that, for high collisional velocities and bodies with sizes comparable to the size of Vesta, we enter into the regime of ``energetic collisions''. If we express the impact energy as the specific kinetic energy of the impactor per unit of the target mass $Q$, following \citet{benz99} we can define a critical threshold for catastrophic disruption of the target body as:

\begin{equation}
Q_D=Q_0\left(\frac{R_{Vesta}}{1 cm}\right)^a+B\rho\left(\frac{R_{Vesta}}{1 cm}\right)^b
\end{equation}

where $R_{Vesta}$ is the radius of Vesta(or target), $\rho$ the density of Vesta (in $g/cm^3$ ). For basalt, for $v = 5$ $km/s$, $B=0.5$ $erg$ $cm^{3}$ $g^{-2}$ , $a=-0.36$, $b=1.36$, $Q_0=9.0\times 10^7$; for $v = 3$ $km/s$, $B=0.3$ $erg$ $cm^{3}$ $g^{-2}$, $a=-0.38$, $b=1.36$, $Q_0=9.0\times 10^7$. 

Analogously to \citet{turrini12}, we considered a collision "energetic" in those cases where $0.05 \cdot Q_D<Q<Q_D$. When $Q>Q_D$, a catastrophic disruption takes place. The mass lost in the last two cases is defined in \citep{benz99}:

\begin{equation}
M_{lost}=M_{Vesta}\left[1+s\left(\frac{Q_0}{Q_D}-1\right)-0.5\right]
\end{equation}

For a basalt target like Vesta and an impact velocity $v = 5$ km/s, $s = 0.35$; for an impact velocity $v = 3 $km/s, $s = 0.5$. It must be noted, however, that such a collision would shatter the target into a cloud of fragments and the largest of them will be about half of the size of the original body. This kind of collisions, therefore, are incompatible with the present structure of Vesta as revealed by Dawn (see \citealt{russell12,consolmagno15}).

For each simulation we compute the mass lost and the mass accreted by Vesta separately for the cases of normal, energetic and catastrophic impacts and we can therefore evaluate the evolution of the mass of Vesta and the percentage of mass lost and/or accreted due to the Late Heavy Bombardment. In order to improve the statistical robustness of our results, we averaged the computed values over $1000$ Monte Carlo extractions of the whole collisional history of Vesta recorded in our dynamical simulations.

\section{Results}\label{sec:Results of 20 Ma evolution}

\subsection{Dynamical results and impact flux on Vesta}

The first test we performed was to compare the depletion factors obtained in our simulations to those obtained by \citet{minton09} and \cite{morbidelli10}. In both cases, we computed the depletion factor considering as our reference initial population of the asteroid belt the one we obtain after the first 5 Ma of our simulations, i.e. immediately before the beginning of the migration process, and comparing it with the one that survives at the end of the simulations. 

In the \citet{minton09} scenario we found a depletion factor of 41\% while in the \citet{morbidelli10} scenario we obtained a depletion factor of 35\%. Before comparing these values with those reported by \citet{minton09} and \citet{morbidelli10}, however, it must be pointed out that our post-migration phase of dynamical clearing (i.e. 5 Ma) is shorter than those considered in the original works. \citet{minton09} report a depletion of 62\% after a 100 Ma temporal interval beginning with the 10 Ma-long migration phase. \citet{morbidelli10} report a depletion of 45\% after the Jumping Jupiters migration phase and the following 25 Ma of dynamical evolution of the asteroid belt (where also the terrestrial planets, Uranus and Neptune were included).

In order to compare our results with the published ones we used the model of the post-LHB evolution of the asteroid belt population by \citet{minton10} to estimate the effects of the missing temporal intervals (85 Ma in the \citealt{minton09} scenario and 20 Ma in the \citealt{morbidelli10} scenario). After these corrective factors were included, we obtained a depletion of 59\% in the \citet{minton09} scenario, comparable to their reported value of 62\%, and 45\% in the \citet{morbidelli10} scenario, again comparable to the reported value of 45\%.

The use of the hybrid symplectic algorithm of MERCURY allowed for an additional test, focusing on the orbital stability of Vesta. As mentioned in Sect. 2.1, each migration scenario was investigated with a set of 32 data-parallel simulations, where the initial condition of Vesta, Jupiter and Saturn were identical among the simulations. Because of its hybrid nature, the adopted algorithm switches between the symplectic method and a numerical solver (specifically, the Bulirsh-Stoer integrator) whenever there is a close encounter between Vesta and a test particle. 

Since the number and the times of the close encounters between the test particles and Vesta will be different in each of the 32 simulations performed for each considered scenario (as each simulation includes test particles populating a different annular region of the asteroid belt, see Sect. 2.1), the dynamical evolution of Vesta will not be exactly identical from a numerical point of view between different simulations. If the orbit of Vesta is intrinsically stable, these small differences between one simulation and another will not affect the overall dynamical behaviour of the asteroid. In the presence of chaos, however, the same small differences could cause the orbital evolution of Vesta to diverge between two simulations of the same set.

We took advantage of this fact to probe the dynamical stability of the orbital region of Vesta by looking for the presence of chaotic effects, i.e. whether these small numerical differences between one simulation and another could result in significant variations of the dynamical evolution of Vesta while starting from the same initial conditions.  
We therefore checked whether Vesta remained in the inner solar system or whether it was removed following its ejection or its collision with one of the other massive bodies: we show the results on Table \ref{tab:20mav} for both the \citet{minton09} scenario and the \citet{morbidelli10} scenario.

\begin{table}[!h]
\begin{center}
\begin{tabular}{|l|c|c|c|}
\hline
\hline   \multicolumn{4}{|c|}{\textbf{Fate of Vesta}} \\
\hline  Scenario & \multicolumn{3}{|c|}{Percentage of bodies}  \\
\hline   & ejected from & collided with & survived \\
   & the Solar System & a massive body &   \\
\hline  \citet{minton09} & $\sim$72\% & $\sim$3\% & 25\%\\
\hline  \citet{morbidelli10} & 0\% & 0\% & 100\%\\
\hline
\hline
\end{tabular}
\caption[Differences in the dynamical evolution of Vesta due to the sensitive dependence on the initial conditions introduced by the hybrid symplectic algorithm in the two scenarios we simulated.]{Differences in the dynamical evolution of Vesta due to the sensitive dependence on the initial conditions introduced by the hybrid symplectic algorithm in the two scenarios we simulated.}
\label{tab:20mav}
\end{center}
\end{table}

We found that in the \citet{minton09} scenario the orbital region of Vesta becomes highly unstable, favouring the ejection of the asteroid from the Solar System: the asteroid is not removed from  our simulations only in 25$\%$ of cases. In the \citet{morbidelli10} scenario, instead, Vesta's orbital region is highly stable and the asteroid is never ejected from the Solar System nor collides with other massive bodies. 
Following the results of this dynamical test, we focused our analysis on the \citet{morbidelli10} scenario, i.e. we choose the planet-planet scattering migration as the reference model.

From the sum of the different impact probabilities $P_{tot}=\sum_{i} {P_{coll}}_i=0.7$ of the $N_{tot}=1.6\times 10^5$ test particles in our simulations we can estimate the number of impacts that should have occurred on Vesta during the LHB. As mentioned above, the initial population of asteroids before the LHB was of the order of $N_{pre-LHB}=7\times 10^6$, therefore the number of impacts, scaled to this population is:

\begin{equation}
N_{imp}(d_{imp}\geq 1 km)=\frac{N_{pre-LHB}}{N_{tot}}P_{tot}=31
\end{equation}

As the uncertainty affecting our results is that characteristic of Poisson statistics, i.e. $\sigma_{n}=\sqrt{n}$, the LHB on Vesta should have produced 31 $\pm$ 5 craters for $d_{imp}\geq 1 km$.
This result can be compared with the number of impacts that would take place in an unperturbed asteroid belt with a constant population 4 times the current one on the same temporal interval, i.e. $\sim 10$, and an unperturbed asteroid belt that rapidly decreases from 4 to 2 times the current population, i.e. $\sim 5$, as can be estimated using the impact probability on Vesta computed by \citet{turrini14}.
Therefore, we find that computed collisions on Vesta during the Late Heavy Bombardment lead to an increase in the number of impacts by about a factor of 5, specifically by a factor $\sim 3$ in the first case and a factor $\sim 6$ in the second case. It is also interesting to compare the number of impacts due to the LHB with the expected number of craters due to asteroids with diameter greater or equal to about 1 km over the last 4 Ga and over the last Ga, i.e. $\sim 1140$ and $\sim 250$ respectively   
\citet{turrini14}. The effect of the LHB is therefore equivalent to 0.12 Ga of collisional evolution of Vesta in the present asteroid belt.

\subsection{Mass evolution of Vesta}

To evaluate the mass lost and accreted by Vesta we used the equations described in Sect. \ref{Mass lost and accreted}, differentiating the cases of normal impacts, energetic impacts and catastrophic impacts.
All the quantities that we estimate are averaged over 1000 Monte Carlo extractions as shown in Tab. \ref{tab:nome}. The eroded mass due to energetic and catastrophic impacts is not considered in the computation of the final mass of Vesta because of their low probability of taking place. The data are divided into the different contributions of normal impacts, energetic impacts and catastrophic impacts. The thickness of the eroded layer is computed using eq. \ref{eqn:vol} assuming that the eroded mass is enclosed in a spherical shell extending outward from the current radius of Vesta $R_V$ \citep{turrini14a}:

\begin{equation}\label{eqn:vol}
r_{eroded}=\left(R_V^3+\frac{3M_{eroded}}{4\pi \rho_V}\right)^{\frac{1}{3}}-R_V
\end{equation}

\begin{table}[!h]
\begin{center}
\begin{tabular}{|l|c|}
\hline
\hline \textbf{NORMAL IMPACTS} & \\
\hline
\hline Number of standard impacts & $31 \pm 5$ \\
\hline 
\hline Eroded mass (g) & $ 1.38\times 10^{19}$ \\
\hline Eroded mass (\%) & $5.33 \times 10^{-3}$\\
\hline Eroded layer (km) & $5.14 \times 10^{-3}$\\
\hline Accreted mass(g) & $5.66\times 10^{18}$ \\
\hline Accreted mass (\%) & $2.19\times 10^{-3}$ \\
\hline Net mass variation & $-8.15\times 10^{18}$ \\
\hline Net eroded mass (\%) & $3.14 \times 10^{-3}$\\
\hline Net eroded layer (km) & $ 3.04  \times 10^{-3}$\\
\hline
\hline \textbf{ENERGETIC IMPACTS} & \\
\hline
\hline Probability of energetic impacts (\%) & $8.9$  \\
\hline
\hline \textbf{CATASTROPHIC IMPACTS} & \\
\hline
\hline Probability of catastrophic impacts (\%) &  $ 1.24\times 10^{-2}$ \\
\hline
\hline
\end{tabular}
\caption[Collisional evolution of Vesta during the LHB (quantities averaged over 1000 Monte Carlo extractions).]{\normalsize{\itshape Collisional evolution of Vesta during the LHB (quantities averaged over 1000 Monte Carlo extractions).}}
\label{tab:nome}
\end{center}
\end{table}
 
The eroded layer during the LHB is only $\sim 3$ m, to be compared to the estimated crustal layer that is $\sim 20-30$ $km$ (\citealt{consolmagno15} and references therein). According to our results, the chances of Vesta undergoing, during the LHB, an energetic impact (like those that created Rheasilvia and Veneneia) is $8.9\%$, while those of a catastrophic impact are $\sim 1.24\times 10^{-2}\%$.
Therefore our results are compatible with Dawn's data and the survival of the crust of Vesta. 

\subsection{Impact craters distribution}

\begin{figure}[!h]
\centering
\includegraphics[width=3in]{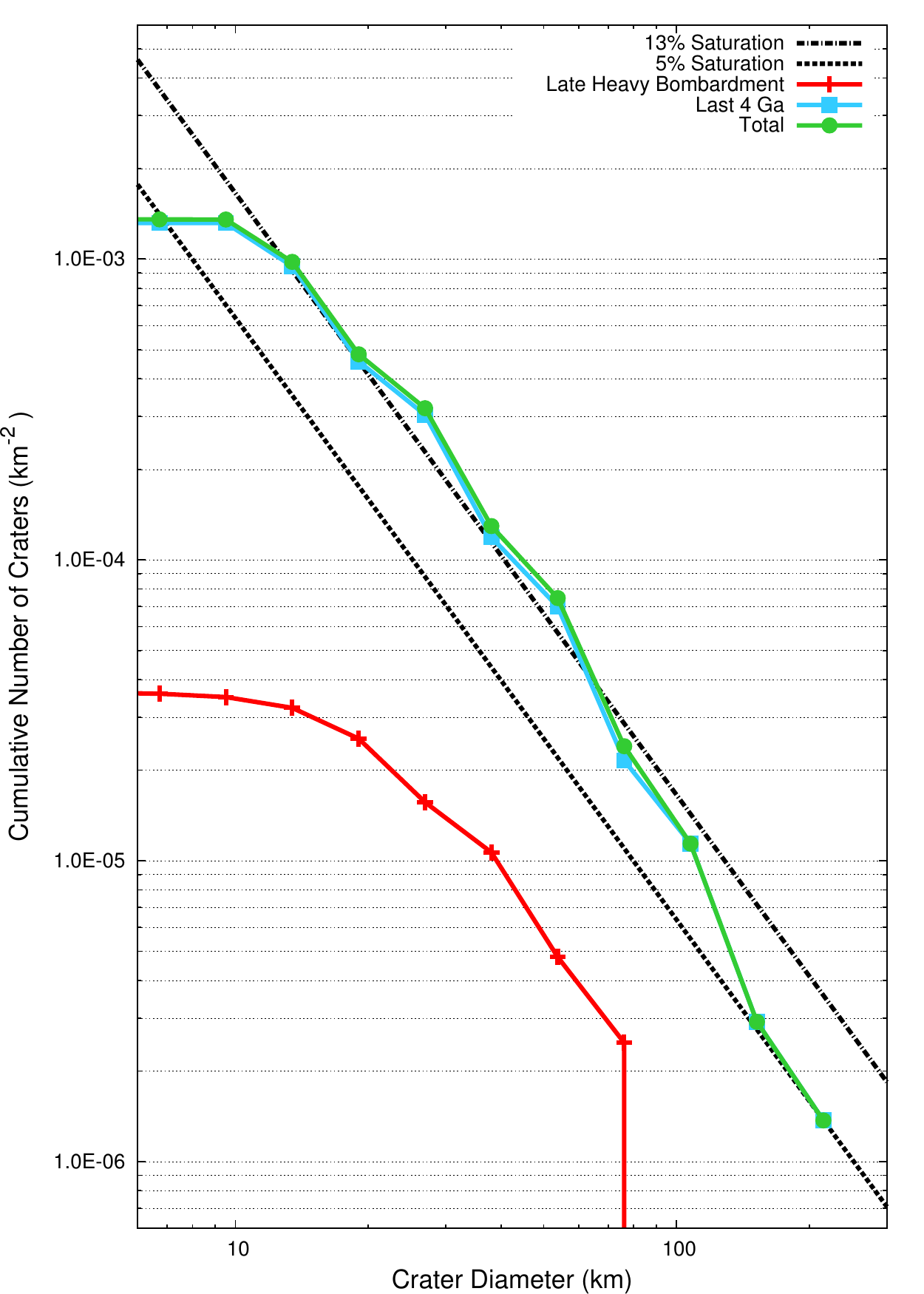}
\caption[Cumulative SFD of the crater population of Vesta produced by the LHB (this work), the last 4 Ga \citep{turrini14} and the sum of the two contributions.]{\itshape{Cumulative SFD of the crater population of Vesta produced by the LHB (this work), the last 4 Ga \citep{turrini14} and the sum of the two contributions. Note that these SFDs do not take into account the effects of  crater erasing processes.}}
\label{fig:lhb-rplot.png}
\end{figure}

In order to compute the effects of the LHB on the cratering of Vesta's surface and to compare them with the cratering resulting from the post-LHB collisional evolution of the asteroid, the crater population produced in our simulations was binned in intervals going from $D$ to $\sqrt{2}D$ with a starting diameter $D=1$ km.
On the ordinate of Fig. \ref{fig:lhb-rplot.png} we show the surface density of the cumulative number of craters (i.e. number of craters per square km, neglecting the effects of crater erasing processes) and on the abscissa we show the cumulative SFD of the crater population of Vesta produced by the LHB, by the last 4 Ga \citep{turrini14} and the sum of the two previous contributions. As a reference, in Fig. \ref{fig:lhb-rplot.png} we also show the values of the crater surface density associated to the 5\% and 13\% levels of geometrical saturation of the surface of Vesta. These values represent respectively the minimum value for which a crater population can reach equilibrium and the value estimated for Mimas, the most cratered surface in the Solar System \citep{melosh89}.

\begin{figure}[!h]
\centering
\includegraphics[width=3in]{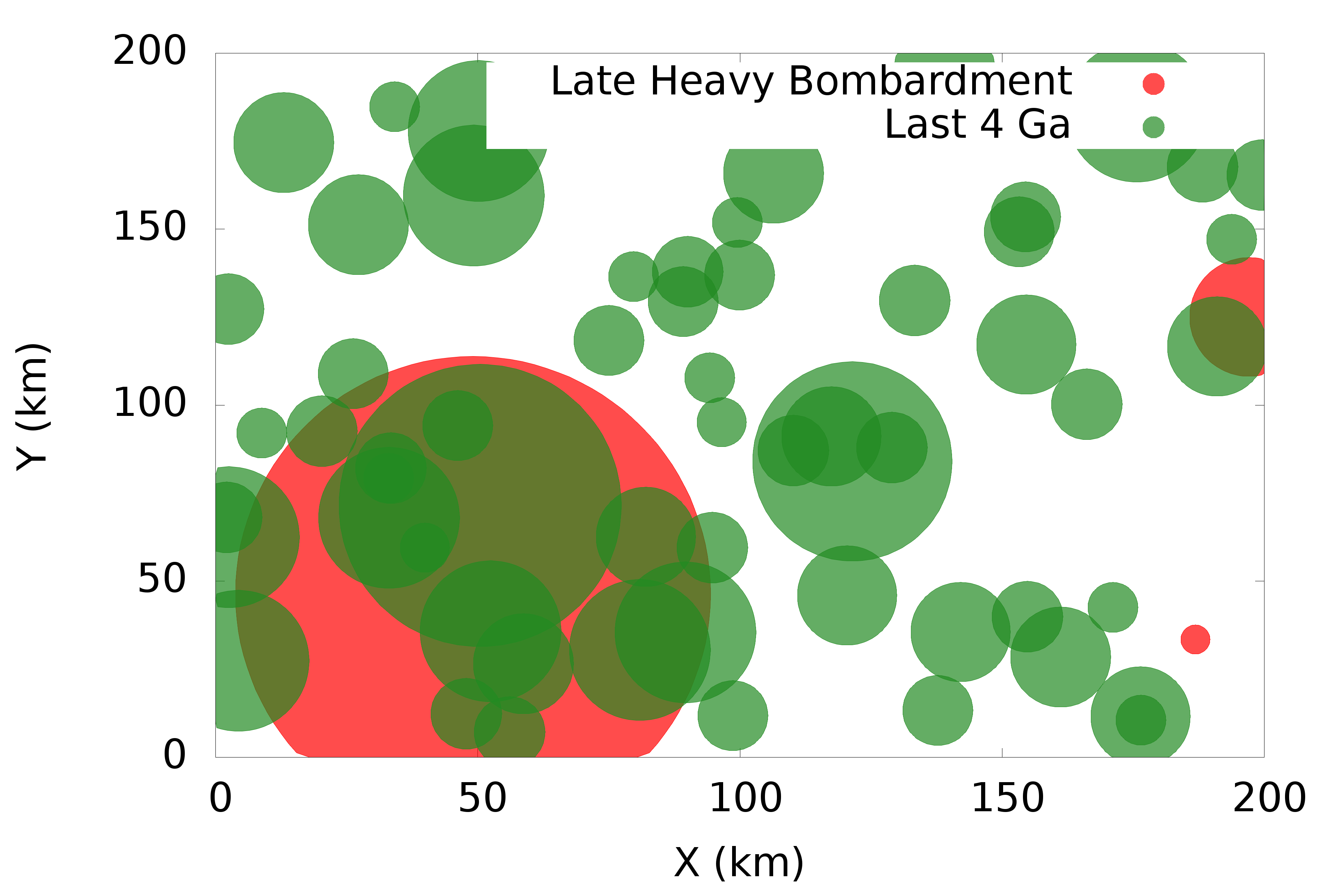}\qquad\qquad
\includegraphics[width=3in]{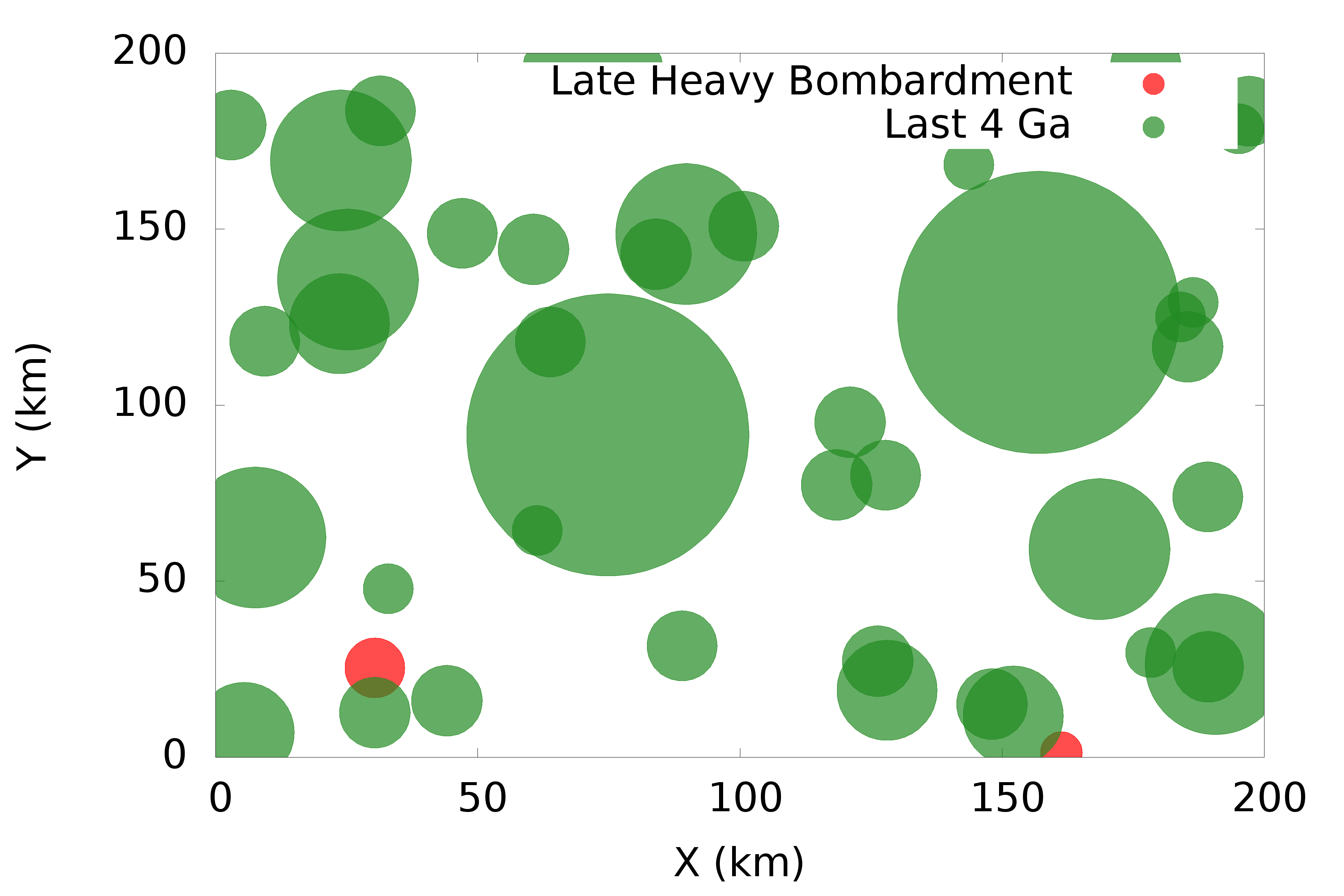}\qquad\qquad
\includegraphics[width=3in]{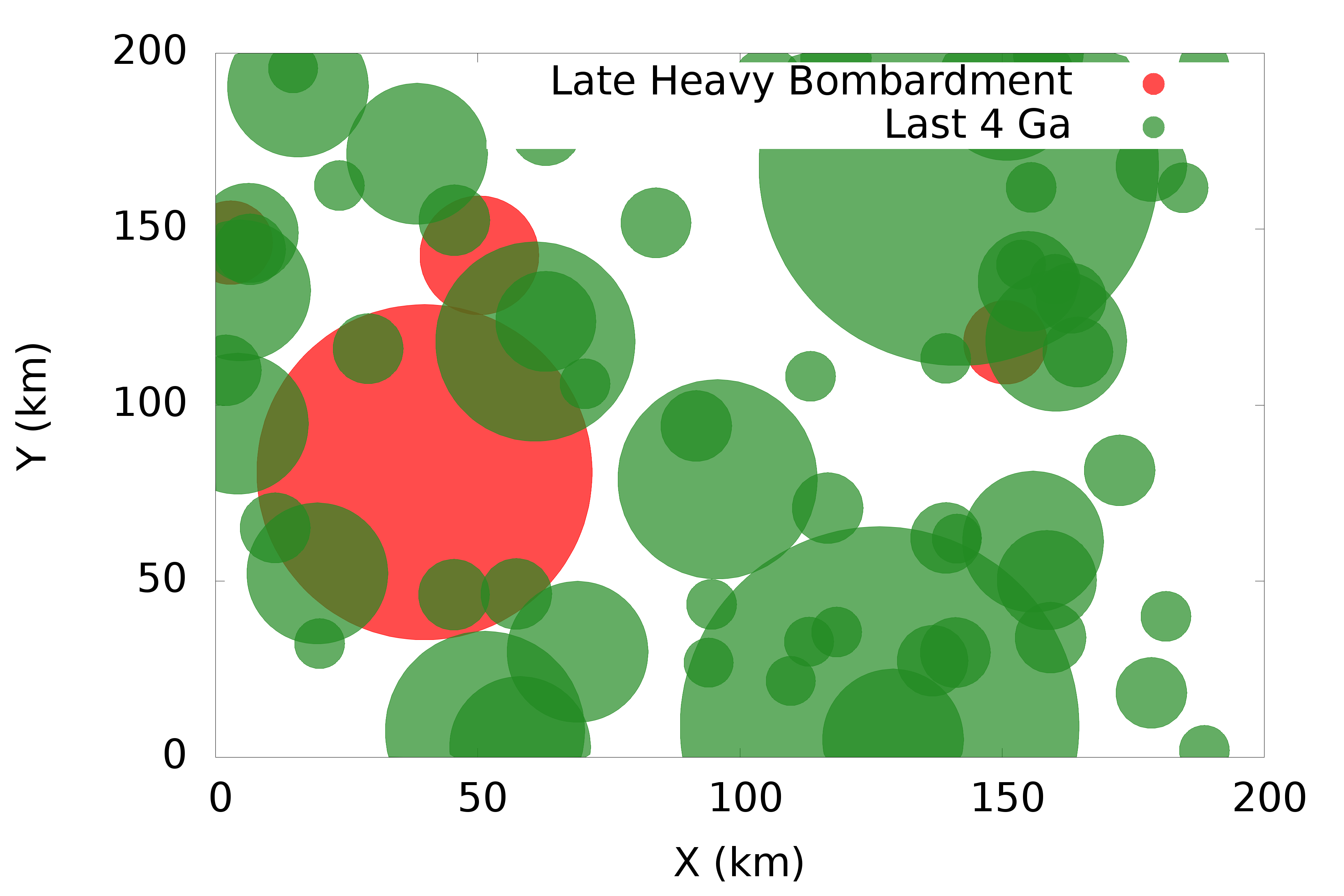}\qquad\qquad
\includegraphics[width=3in]{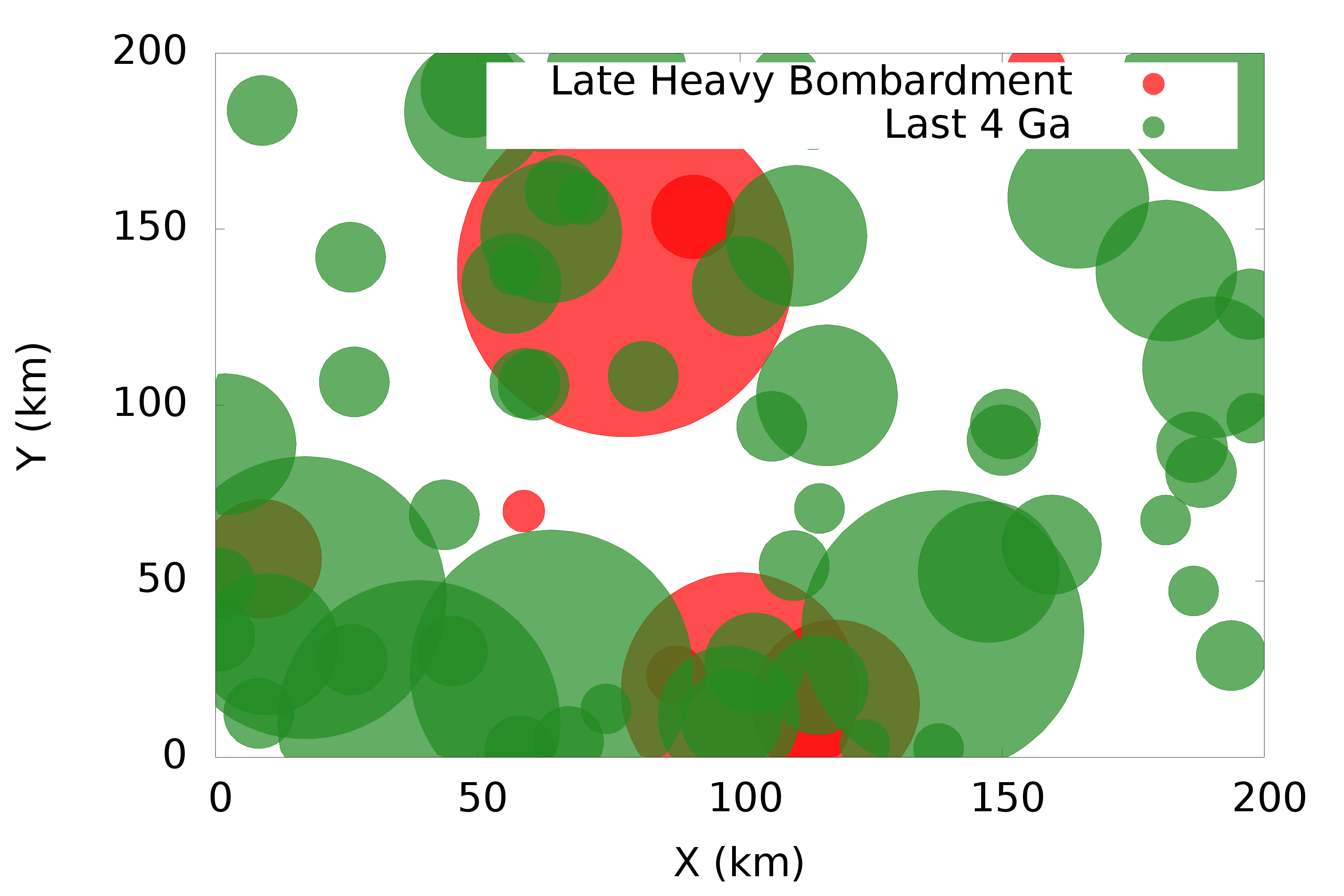}
\caption[Four examples of crater population due produced by the LHB and by the last 4 Ga on a surface of $200\times200$ km of Vesta.]{\itshape{Four examples of crater population due produced by the LHB and by the last 4 Ga on a surface of $200\times200$ km of Vesta.}}
\label{fig:1234}
\end{figure}

From Fig. \ref{fig:lhb-rplot.png} we can immediately see that the contribution of the LHB to the crater population of Vesta is quite limited with respect to that of the post-LHB collisional evolution of the asteroid, and that the contribution of the LHB to the saturation of the surface of Vesta is negligible. The comparison between the contribution of the LHB and that of the last 4 Ga also reveals that the post-LHB collisional evolution of Vesta most plausibly cancelled all signatures of the LHB itself. Fig. \ref{fig:1234} shows a four examples of crater populations (estimated by means of Monte Carlo extractions) produced by the LHB and by the last 4 Ga on a region of interest of $200\times200$ km on Vesta. As can be seen in all cases, the older LHB craters are always partially or totally erased by the younger craters, making their identification difficult, if not impossible. According to our results it would be possible, in principle, to investigate the pre-LHB craters only with large basins ($ \geq 200 km$). However, the current data from Dawn mission do not show the existence of such large basins that are unequivocally older than the LHB \citep{schenk12,obrien14,schmedemann14}. 

\section{Discussion and conclusions}

Our study aimed to assess the effects of the LHB in the Nice Model \citep{tsiganis05,gomes05,morbidelli05} on Vesta. The Nice Model explains the Late Heavy Bombardment as the result of the migration of the giant planets in the outer Solar System, which we modelled with the semi-analytic approach proposed by \citet{minton09}. Vesta plays an important role in the study of the evolution of the Solar System because it is the only asteroid we know that formed and differentiated in the first few million years of the life of Solar System, that survived intact until now and of which we possess samples in the form of HEDs meteorites. In particular, the survival of the relatively thin basaltic crust put strong constraints to the collisional evolution of the asteroid during the life of the Solar System. Using the collisional model developed by \citet{turrini11,turrini12} to study the pre-LHB evolution of Vesta we simulated the impact history of the asteroid during the Late Heavy Bombardment.

Our results showed that the orbital region of Vesta during the planetesimal-driven migration studied by \citet{minton09} is dominated by chaotic effects, resulting highly unstable. Because of the small numerical differences in the orbital evolution of Vesta introduced by the hybrid symplectic algorithm (which are due to the different number and times of the close encounters between test particles and Vesta in each of the 32 simulations used to investigate the scenario), the asteroid survived from being removed from the Solar System only in 25\% of cases. Even in those cases where it survived, moreover, the orbit of Vesta at the end of our simulations was far more eccentric than the present one ($e\approx0.4$ vs $e\approx0.09$). In those cases where it is removed from the belt, Vesta is ejected from the Solar System in 95.8\% of the cases and collides with the Sun in 4.2\% of the cases. These events occur after 6-7 Ma from the beginning of our integrations, i.e.  after the giant planets completed more than 90-99\% of their migration. On the contrary, in our set of 32 simulations focusing on the planet-planet scattering migration described in the \citet{morbidelli10} scenario, we did not find significant qualitative variations of the dynamical evolution of Vesta between one simulation and the others due to the small numerical differences introduced by the hybrid symplectic algorithm.  The asteroid survived and had the same orbital evolution in all 32 simulations, which indicates that the \textcolor{black}{orbital region} of Vesta remains intrinsically stable during the LHB and proves the planet-planet scattering migration a more favourable scenario for the survival of Vesta. Taking into account these results, we focused our analysis on the collisional evolution of Vesta in the \citet{morbidelli10} scenario.

From our results we estimated the probability of Vesta undergoing energetic and/or catastrophic impacts during the LHB, finding  that these two types of collisions are reasonably rare. Energetic and catastrophic impacts from asteroidal impactors have a probability of occurrence of $8.9\%$ and $1.24\times 10^{-2}\%$ respectively, far lower than the number of such events as estimated by \citet{broz13} for cometary impactors during the LHB, i.e. $2\pm1.4$. The estimated number of impacts of asteroids with diameter larger than or equal to 1 km with Vesta during the LHB is $31\pm5$, a factor $\sim5$ larger than it would be expected in an unperturbed asteroid belt during the same time interval. The contribution of the putative extended belt \citep{bottke12} to the collisional evolution of Vesta during the LHB is quite limited, being of about $2\pm1$ impact. We used our data to estimate the erosion of the surface of Vesta during the LHB, to test whether our results are consistent with the survival of its basaltic crust. Averaging on $1000$ Monte Carlo extractions, we computed a net mass loss of $ \sim 8.15\times 10^{18}$ g, i.e.  $0.003 \% $ of the present mass of Vesta, equivalent to the erosion of a layer with thickness of about 3 m. It must be pointed out that this net mass loss results from the balance between the erosion of the vestan surface and its contamination by exogenous material, i.e. the mass loss from the basaltic crust and the mass gain mainly in the form of chondritic material (see also \citealt{turrini14} for a discussion). If we focus only on the erosion of the basaltic material, ignoring the contribution of contamination to the mass balance, the LHB results in the loss of a layer with thickness of 5 m. 

When compared to the post-LHB collisional evolution of Vesta, our data indicate that the contribution of the LHB to the saturation of the surface of Vesta in the Nice Model scenario is not significant. On the contrary the effects of the last $\sim4$ Ga, already resulting in the saturation of the surface of Vesta to a level greater than 5$\%$ for craters smaller than $200$ km in diameter, efficiently erase the crater population produced by the LHB in agreement with the fact that the crater chronologies obtained by the Dawn mission do not show any region of the surface of Vesta unequivocally identifiable as older than 4.0 Ga \citep{marchi12,obrien14,schmedemann14}. Nevertheless, the results obtained in our studies indicate an effective increase in the number of collisions on Vesta as a result of migration of the giant planets, quantifiable in about a factor of 5 with respect to an unperturbed asteroid belt as mentioned previously. Globally, the mass loss, the eroded surface layer and the cratering of Vesta during the LHB we estimated based on our simulations are compatible with the current morphology and mineralogy of Vesta as revealed by the Dawn mission and the HED meteorites. Our results lead us to conclude that the planet-planet scattering migration mechanism, like that discussed by \citet{morbidelli10} and references therein, is compatible with the collisional and dynamical survival of Vesta to the Late Heavy Bombardment and with the current knowledge of the characteristics of Vesta.

\section*{Acknowledgements}

The authors wish to thank Alessandro Morbidelli, David O'Brien and an anonymous referee for their comments that helped improve this work. This research has been partially supported by the Italian Space Agency (ASI) and by the International Space Science Institute (ISSI) in Bern through the International Teams 2012 project ``Vesta, the key to the origins of the Solar System''. The computational resources used in this research have been supplied by INAF-IAPS through the projects ``DataWell'' and ``HPP -- High Performance Planetology''.

\bibliography{Pirani_Icarus}{}

\begin{thebibliography}{40}
\expandafter\ifx\csname natexlab\endcsname\relax\def\natexlab#1{#1}\fi
\expandafter\ifx\csname url\endcsname\relax
  \def\url#1{\texttt{#1}}\fi
\expandafter\ifx\csname urlprefix\endcsname\relax\def\urlprefix{URL }\fi

\bibitem[{{Benz} and {Asphaug}(1999)}]{benz99}
{Benz}, W., {Asphaug}, E., Nov. 1999. {Catastrophic Disruptions Revisited}.
  Icarus 142, 5--20.

\bibitem[{{Bizzarro} et~al.(2005){Bizzarro}, {Baker}, {Haack}, and
  {Lundgaard}}]{bizzarro05}
{Bizzarro}, M., {Baker}, J.~A., {Haack}, H., {Lundgaard}, K.~L., Oct. 2005.
  {Rapid Timescales for Accretion and Melting of Differentiated Planetesimals
  Inferred from $^{26}$Al-$^{26}$Mg Chronometry}. The Astrophysical Journal
  Letters 632, L41--L44.

\bibitem[{{Bottke} et~al.(2005){Bottke}, {Durda}, {Nesvorn{\'y}}, {Jedicke},
  {Morbidelli}, {Vokrouhlick{\'y}}, and {Levison}}]{bottke05}
{Bottke}, W.~F., {Durda}, D.~D., {Nesvorn{\'y}}, D., {Jedicke}, R.,
  {Morbidelli}, A., {Vokrouhlick{\'y}}, D., {Levison}, H., May 2005. {The
  fossilized size distribution of the main asteroid belt}. Icarus 175,
  111--140.

\bibitem[{{Bottke} et~al.(2012){Bottke}, {Vokrouhlick{\'y}}, {Minton},
  {Nesvorn{\'y}}, {Morbidelli}, {Brasser}, {Simonson}, and
  {Levison}}]{bottke12}
{Bottke}, W.~F., {Vokrouhlick{\'y}}, D., {Minton}, D., {Nesvorn{\'y}}, D.,
  {Morbidelli}, A., {Brasser}, R., {Simonson}, B., {Levison}, H.~F., May 2012.
  {An Archaean heavy bombardment from a destabilized extension of the asteroid
  belt}. Nature 485, 78--81.

\bibitem[{{Britt} et~al.(2002){Britt}, {Yeomans}, {Housen}, and
  {Consolmagno}}]{britt02}
{Britt}, D.~T., {Yeomans}, D., {Housen}, K., {Consolmagno}, G., 2002. {Asteroid
  Density, Porosity, and Structure}. Asteroids III, 485--500.

\bibitem[{{Bro{\v z}} et~al.(2013){Bro{\v z}}, {Morbidelli}, {Bottke},
  {Rozehnal}, {Vokrouhlick{\'y}}, and {Nesvorn{\'y}}}]{broz13}
{Bro{\v z}}, M., {Morbidelli}, A., {Bottke}, W.~F., {Rozehnal}, J.,
  {Vokrouhlick{\'y}}, D., {Nesvorn{\'y}}, D., Mar. 2013. {Constraining the
  cometary flux through the asteroid belt during the late heavy bombardment}.
  Astronomy \& Astrophysics 551, A117.

\bibitem[{{Chambers}(1999)}]{chambers99}
{Chambers}, J.~E., apr 1999. A hybrid symplectic integrator that permits close
  encounters between massive bodies. Monthly Notices of the Royal Astronomical
  Society 304, 793--799.

\bibitem[{Consolmagno et~al.(2015)Consolmagno, Golabek, Turrini, Jutzi, Sirono,
  Svetsov, and Tsiganis}]{consolmagno15}
Consolmagno, G., Golabek, G., Turrini, D., Jutzi, M., Sirono, S., Svetsov, V.,
  Tsiganis, K., 2015. Is vesta an intact and pristine protoplanet? Icarus
  254~(0), 190 -- 201.
\newline\urlprefix\url{http://www.sciencedirect.com/science/article/pii/S001910351500130X}

\bibitem[{{Consolmagno} and {Drake}(1977)}]{consolmagno77}
{Consolmagno}, G.~J., {Drake}, M.~J., Sep. 1977. {Composition and evolution of
  the eucrite parent body - Evidence from rare earth elements}. Geochimica et
  Cosmochimica Acta 41, 1271--1282.

\bibitem[{{De Sanctis} et~al.(2012){De Sanctis}, {Ammannito}, {Capria}, {Tosi},
  {Capaccioni}, {Zambon}, {Carraro}, {Fonte}, {Frigeri}, {Jaumann}, {Magni},
  {Marchi}, {McCord}, {McFadden}, {McSween}, {Mittlefehldt}, {Nathues},
  {Palomba}, {Pieters}, {Raymond}, {Russell}, {Toplis}, and
  {Turrini}}]{desanctis12}
{De Sanctis}, M.~C., {Ammannito}, E., {Capria}, M.~T., {Tosi}, F.,
  {Capaccioni}, F., {Zambon}, F., {Carraro}, F., {Fonte}, S., {Frigeri}, A.,
  {Jaumann}, R., {Magni}, G., {Marchi}, S., {McCord}, T.~B., {McFadden}, L.~A.,
  {McSween}, H.~Y., {Mittlefehldt}, D.~W., {Nathues}, A., {Palomba}, E.,
  {Pieters}, C.~M., {Raymond}, C.~A., {Russell}, C.~T., {Toplis}, M.~J.,
  {Turrini}, D., May 2012. {Spectroscopic Characterization of Mineralogy and
  Its Diversity Across Vesta}. Science 336, 697--700.

\bibitem[{{Duncan} et~al.(1998){Duncan}, {Levison}, and {Lee}}]{duncan98}
{Duncan}, M.~J., {Levison}, H.~F., {Lee}, M.~H., Oct. 1998. {A Multiple Time
  Step Symplectic Algorithm for Integrating Close Encounters}. The Astronomical
  Journal 116, 2067--2077.

\bibitem[{Gaffey(1997)}]{gaffey97}
Gaffey, M.~J., 1997. Surface lithologic heterogeneity of asteroid 4 vesta.
  Icarus 127~(1), 130 -- 157.
\newline\urlprefix\url{http://www.sciencedirect.com/science/article/pii/S0019103597956803}

\bibitem[{Gomes et~al.(2005)Gomes, Levison, Tsiganis, and Morbidelli}]{gomes05}
Gomes, R., Levison, H.~F., Tsiganis, K., Morbidelli, A., 2005. {Origin of the
  cataclysmic Late Heavy Bombardment period of the terrestrial planets}. Nature
  435~(7041), 466--469.
\newline\urlprefix\url{http://dx.doi.org/10.1038/nature03676}

\bibitem[{Holsapple and Housen(2007)}]{holsapple07}
Holsapple, K.~A., Housen, K.~R., 2007. A crater and its ejecta: An
  interpretation of deep impact. Icarus 191~(2, Supplement), 586 -- 597.
\newline\urlprefix\url{http://www.sciencedirect.com/science/article/pii/S0019103507004368}

\bibitem[{{Jedicke} et~al.(2002){Jedicke}, {Larsen}, and {Spahr}}]{Jedicke02}
{Jedicke}, R., {Larsen}, J., {Spahr}, T., 2002. {Observational Selection
  Effects in Asteroid Surveys}. Asteroids III, 71--87.

\bibitem[{{Levison} et~al.(2011){Levison}, {Walsh}, {Barr}, and
  {Dones}}]{Levison11}
{Levison}, H.~F., {Walsh}, K.~J., {Barr}, A.~C., {Dones}, L., Aug. 2011. {Ridge
  formation and de-spinning of Iapetus via an impact-generated satellite}.
  Icarus 214, 773--778.

\bibitem[{{Marchi} et~al.(2014){Marchi}, {Bottke}, {O'Brien}, {Schenk},
  {Mottola}, {De Sanctis}, {Kring}, {Williams}, {Raymond}, and
  {Russell}}]{Marchi13}
{Marchi}, S., {Bottke}, W.~F., {O'Brien}, D.~P., {Schenk}, P., {Mottola}, S.,
  {De Sanctis}, M.~C., {Kring}, D.~A., {Williams}, D.~A., {Raymond}, C.~A.,
  {Russell}, C.~T., Nov. 2014. {Small crater populations on Vesta}. Planetary
  Space Science 103, 96--103.

\bibitem[{{Marchi} et~al.(2012){Marchi}, {McSween}, {O'Brien}, {Schenk}, {De
  Sanctis}, {Gaskell}, {Jaumann}, {Mottola}, {Preusker}, {Raymond}, {Roatsch},
  and {Russell}}]{marchi12}
{Marchi}, S., {McSween}, H.~Y., {O'Brien}, D.~P., {Schenk}, P., {De Sanctis},
  M.~C., {Gaskell}, R., {Jaumann}, R., {Mottola}, S., {Preusker}, F.,
  {Raymond}, C.~A., {Roatsch}, T., {Russell}, C.~T., May 2012. {The Violent
  Collisional History of Asteroid 4 Vesta}. Science 336, 690--.

\bibitem[{{McCord} et~al.(1970){McCord}, {Adams}, and {Johnson}}]{mccord70}
{McCord}, T.~B., {Adams}, J.~B., {Johnson}, T.~V., Jun. 1970. {Asteroid Vesta:
  Spectral Reflectivity and Compositional Implications}. Science 168,
  1445--1447.

\bibitem[{McSween et~al.(2011)McSween, Mittlefehldt, Beck, Mayne, and
  McCoy}]{mcsween11}
McSween, HarryY., J., Mittlefehldt, D., Beck, A., Mayne, R., McCoy, T., 2011.
  Hed meteorites and their relationship to the geology of vesta and the dawn
  mission. Space Science Reviews 163~(1-4), 141--174.
\newline\urlprefix\url{http://dx.doi.org/10.1007/s11214-010-9637-z}

\bibitem[{Melosh(1989)}]{melosh89}
Melosh, H., 1989. Impact cratering: a geologic process. Oxford monographs on
  geology and geophysics. Oxford University Press.
\newline\urlprefix\url{https://books.google.se/books?id=nZwRAQAAIAAJ}

\bibitem[{{Minton} and {Malhotra}(2009)}]{minton09}
{Minton}, D.~A., {Malhotra}, R., Feb. 2009. {A record of planet migration in
  the main asteroid belt}. Nature 457, 1109--1111.

\bibitem[{{Minton} and {Malhotra}(2010)}]{minton10}
{Minton}, D.~A., {Malhotra}, R., Jun. 2010. {Dynamical erosion of the asteroid
  belt and implications for large impacts in the inner Solar System}. Icarus
  207, 744--757.

\bibitem[{{Morbidelli} et~al.(2010){Morbidelli}, {Brasser}, {Gomes}, {Levison},
  and {Tsiganis}}]{morbidelli10}
{Morbidelli}, A., {Brasser}, R., {Gomes}, R., {Levison}, H.~F., {Tsiganis}, K.,
  Nov. 2010. {Evidence from the Asteroid Belt for a Violent Past Evolution of
  Jupiter's Orbit}. The Astronomical Journal 140, 1391--1401.

\bibitem[{Morbidelli et~al.(2005)Morbidelli, Levison, Tsiganis, and
  Gomes}]{morbidelli05}
Morbidelli, A., Levison, H.~F., Tsiganis, K., Gomes, R., 2005. {Chaotic capture
  of Jupiter's Trojan asteroids in the early Solar System}. Nature 435~(7041),
  462--465.
\newline\urlprefix\url{http://dx.doi.org/10.1038/nature03540}

\bibitem[{{Morbidelli} et~al.(2007){Morbidelli}, {Tsiganis}, {Crida},
  {Levison}, and {Gomes}}]{morbidelli07}
{Morbidelli}, A., {Tsiganis}, K., {Crida}, A., {Levison}, H.~F., {Gomes}, R.,
  Nov. 2007. {Dynamics of the Giant Planets of the Solar System in the Gaseous
  Protoplanetary Disk and Their Relationship to the Current Orbital
  Architecture}. Astronomical Journal 134, 1790--1798.

\bibitem[{{O'Brien} et~al.(2014){O'Brien}, {Marchi}, {Morbidelli}, {Bottke},
  {Schenk}, {Russell}, and {Raymond}}]{obrien14}
{O'Brien}, D.~P., {Marchi}, S., {Morbidelli}, A., {Bottke}, W.~F., {Schenk},
  P.~M., {Russell}, C.~T., {Raymond}, C.~A., Nov. 2014. {Constraining the
  cratering chronology of Vesta}. Planetary and Space Science 103, 131--142.

\bibitem[{Prettyman et~al.(2012)Prettyman, Mittlefehldt, Yamashita, Lawrence,
  Beck, Feldman, McCoy, McSween, Toplis, Titus, Tricarico, Reedy, Hendricks,
  Forni, Le~Corre, Li, Mizzon, Reddy, Raymond, and Russell}]{prettyman12}
Prettyman, T.~H., Mittlefehldt, D.~W., Yamashita, N., Lawrence, D.~J., Beck,
  A.~W., Feldman, W.~C., McCoy, T.~J., McSween, H.~Y., Toplis, M.~J., Titus,
  T.~N., Tricarico, P., Reedy, R.~C., Hendricks, J.~S., Forni, O., Le~Corre,
  L., Li, J.-Y., Mizzon, H., Reddy, V., Raymond, C.~A., Russell, C.~T., 2012.
  Elemental mapping by dawn reveals exogenic h in vesta’s regolith. Science
  338~(6104), 242--246.
\newline\urlprefix\url{http://www.sciencemag.org/content/338/6104/242.abstract}

\bibitem[{{Russell} et~al.(2012){Russell}, {Raymond}, {Coradini}, {McSween},
  {Zuber}, {Nathues}, {De Sanctis}, {Jaumann}, {Konopliv}, {Preusker}, {Asmar},
  {Park}, {Gaskell}, {Keller}, {Mottola}, {Roatsch}, {Scully}, {Smith},
  {Tricarico}, {Toplis}, {Christensen}, {Feldman}, {Lawrence}, {McCoy},
  {Prettyman}, {Reedy}, {Sykes}, and {Titus}}]{russell12}
{Russell}, C.~T., {Raymond}, C.~A., {Coradini}, A., {McSween}, H.~Y., {Zuber},
  M.~T., {Nathues}, A., {De Sanctis}, M.~C., {Jaumann}, R., {Konopliv}, A.~S.,
  {Preusker}, F., {Asmar}, S.~W., {Park}, R.~S., {Gaskell}, R., {Keller},
  H.~U., {Mottola}, S., {Roatsch}, T., {Scully}, J.~E.~C., {Smith}, D.~E.,
  {Tricarico}, P., {Toplis}, M.~J., {Christensen}, U.~R., {Feldman}, W.~C.,
  {Lawrence}, D.~J., {McCoy}, T.~J., {Prettyman}, T.~H., {Reedy}, R.~C.,
  {Sykes}, M.~E., {Titus}, T.~N., May 2012. {Dawn at Vesta: Testing the
  Protoplanetary Paradigm}. Science 336, 684--.

\bibitem[{Schenk et~al.(2012)Schenk, O'Brien, Marchi, Gaskell, Preusker,
  Roatsch, Jaumann, Buczkowski, McCord, McSween, Williams, Yingst, Raymond, and
  Russell}]{schenk12}
Schenk, P., O'Brien, D.~P., Marchi, S., Gaskell, R., Preusker, F., Roatsch, T.,
  Jaumann, R., Buczkowski, D., McCord, T., McSween, H.~Y., Williams, D.,
  Yingst, A., Raymond, C., Russell, C., 2012. The geologically recent giant
  impact basins at vesta's south pole. Science 336~(6082), 694--697.
\newline\urlprefix\url{http://www.sciencemag.org/content/336/6082/694.abstract}

\bibitem[{{Schiller} et~al.(2011){Schiller}, {Baker}, {Creech}, {Paton},
  {Millet}, {Irving}, and {Bizzarro}}]{schiller11}
{Schiller}, M., {Baker}, J., {Creech}, J., {Paton}, C., {Millet}, M.-A.,
  {Irving}, A., {Bizzarro}, M., Oct. 2011. {Rapid Timescales for Magma Ocean
  Crystallization on the Howardite-Eucrite-Diogenite Parent Body}. The
  Astrophysical Journal Letters 740, L22.

\bibitem[{{Schmedemann} et~al.(2014){Schmedemann}, {Kneissl}, {Ivanov},
  {Michael}, {Wagner}, {Neukum}, {Ruesch}, {Hiesinger}, {Krohn}, {Roatsch},
  {Preusker}, {Sierks}, {Jaumann}, {Reddy}, {Nathues}, {Walter}, {Neesemann},
  {Raymond}, and {Russell}}]{schmedemann14}
{Schmedemann}, N., {Kneissl}, T., {Ivanov}, B.~A., {Michael}, G.~G., {Wagner},
  R.~J., {Neukum}, G., {Ruesch}, O., {Hiesinger}, H., {Krohn}, K., {Roatsch},
  T., {Preusker}, F., {Sierks}, H., {Jaumann}, R., {Reddy}, V., {Nathues}, A.,
  {Walter}, S.~H.~G., {Neesemann}, A., {Raymond}, C.~A., {Russell}, C.~T., Nov.
  2014. {The cratering record, chronology and surface ages of (4) Vesta in
  comparison to smaller asteroids and the ages of HED meteorites}. Planetary
  and Space Science 103, 104--130.

\bibitem[{{Svetsov}(2011)}]{svetsov11}
{Svetsov}, V., Jul. 2011. {Cratering erosion of planetary embryos}. Icarus 214,
  316--326.

\bibitem[{Tsiganis et~al.(2005)Tsiganis, Gomes, Morbidelli, and
  Levison}]{tsiganis05}
Tsiganis, K., Gomes, R., Morbidelli, A., Levison, H.~F., May 2005. {Origin of
  the orbital architecture of the giant planets of the Solar System}. Nature
  435~(7041), 459--461.
\newline\urlprefix\url{http://dx.doi.org/10.1038/nature03539}

\bibitem[{{Turrini}(2014)}]{turrini14a}
{Turrini}, D., Nov. 2014. {The primordial collisional history of Vesta: crater
  saturation, surface evolution and survival of the basaltic crust}. Planetary
  and Space Science 103, 82--95.

\bibitem[{{Turrini} et~al.(2014){Turrini}, {Combe}, {McCord}, {Oklay},
  {Vincent}, {Prettyman}, {McSween}, {Consolmagno}, {De Sanctis}, {Le Corre},
  {Longobardo}, {Palomba}, and {Russell}}]{turrini14}
{Turrini}, D., {Combe}, J.-P., {McCord}, T.~B., {Oklay}, N., {Vincent}, J.-B.,
  {Prettyman}, T.~H., {McSween}, H.~Y., {Consolmagno}, G.~J., {De Sanctis},
  M.~C., {Le Corre}, L., {Longobardo}, A., {Palomba}, E., {Russell}, C.~T.,
  Sep. 2014. {The contamination of the surface of Vesta by impacts and the
  delivery of the dark material}. Icarus 240, 86--102.

\bibitem[{Turrini et~al.(2012)Turrini, Coradini, and Magni}]{turrini12}
Turrini, D., Coradini, A., Magni, G., 2012. Jovian early bombardment:
  Planetesimal erosion in the inner asteroid belt. The Astrophysical Journal
  750~(1), 8.
\newline\urlprefix\url{http://stacks.iop.org/0004-637X/750/i=1/a=8}

\bibitem[{Turrini et~al.(2011)Turrini, Magni, and Coradini}]{turrini11}
Turrini, D., Magni, G., Coradini, A., jan 2011. Probing the history of solar
  system through the cratering records on vesta and ceres. Monthly Notices of
  the Royal Astronomical Society 413, 2439--2466.

\bibitem[{{Turrini} and {Svetsov}(2014)}]{turrini14b}
{Turrini}, D., {Svetsov}, V., Jan. 2014. {The Formation of Jupiter, the Jovian
  Early Bombardment and the Delivery of Water to the Asteroid Belt: The Case of
  (4) Vesta}. Life 4, 4--34.

\bibitem[{{Wisdom} and {Holman}(1991)}]{wisdom91}
{Wisdom}, J., {Holman}, M., Oct. 1991. {Symplectic maps for the n-body
  problem}. The Astronomical Journal 102, 1528--1538.

\end{thebibliography}
\bibliographystyle{elsarticle-harv}

\end{document}